\documentstyle[12pt,epsfig]{article}
\def\be{\begin{equation}}
\def\ee{\end{equation}}
\def\bea{\begin{eqnarray}}
\def\eea{\end{eqnarray}}
\def\ba{\begin{array}}
\def\ea{\end{array}}
\def\calo{{\cal O}}
\def\gappeq{\mathrel{\rlap {\raise.5ex\hbox{$>$}}
{\lower.5ex\hbox{$\sim$}}}}
\def\permil{$\%\raise.20ex\hbox{$_0$}}
\def\lappeq{\mathrel{\rlap{\raise.5ex\hbox{$<$}}
{\lower.5ex\hbox{$\sim$}}}}

\begin{document}
\topmargin -1.0cm
\oddsidemargin -0.8cm
\evensidemargin -0.8cm
\pagestyle{empty}
\begin{flushright}
UAB-FT-413\\
February, 1997
\end{flushright}
\vspace*{5mm}
\vspace{5mm}
\begin{center}
{\Large\bf Dynamical determination}\\
\vspace{0.5cm}
{\Large\bf  of the Supersymmetric Higgs mass}\\
\vspace{2cm}
{\large\bf Paolo Ciafaloni and Alex Pomarol}\\
\vspace{.4cm}
{Institut de F{\'\i}sica d'Altes Energies}\\
{Universitat Aut{\`o}noma de Barcelona}\\  
{E-08193 Bellaterra, Barcelona, Spain}\\
\end{center}
\vspace{2cm}
\begin{abstract}

Considering  the supersymmetric Higgs mass ($\mu$-parameter)
as a dynamical variable to be determined by minimizing
the energy, we predict its value as a function of the 
soft masses of the potential.
We find that  $\mu$ has a nonzero value  close to the weak scale.
This scenario offers  a simultaneous
 solution to the
doublet-triplet splitting problem and to the $\mu$-problem. 
We discuss its viability in theories with gauge mediated 
supersymmetry breaking.

\end{abstract}

\vfill
\eject
\pagestyle{empty}
\setcounter{page}{1}
\setcounter{footnote}{0}
\pagestyle{plain}


\section{Introduction}

In the supersymmetric standard model  (MSSM), 
the  Higgs doublet superfields   
 present three important features that distinguish
them from the lepton and  quark superfields:

\noindent a) 
The Higgs superfield, $H+\bar H$, is
 vector-like  under the standard model (SM) group
  and therefore it is  allowed to have
a large supersymmetric mass $\mu H\bar H$. 

\noindent b) If we grand unify the MSSM in a theory such  as SU(5), 
the Higgs doublets cannot be  embedded in a complete 
GUT-representation.

\noindent c) The  scalar components of the Higgs doublets 
 have to get  nonzero vacuum expectation
values (VEVs) to break the electroweak symmetry.
 
Properties (a)  and (c)  lead to the $\mu$-problem. 
If the Higgs doublets have to get  nonzero VEVs, 
the value of $\mu$ has to be
bounded  from above by  the  weak scale.
On the other hand,
Higgsino searches at LEP1.5 \cite{aleph} put a lower 
bound on $\mu$   roughly given by $|\mu|\gappeq 50$ GeV.
Due to property (a), there is, a priori, no reason to expect the 
value of $\mu$  to be in this small window; 
this is referred as the 
 $\mu$-problem. This problem is especially severe in theories with
gauge mediated supersymmetry breaking (GMSB) \cite{us}.
In these theories the supersymmetry breaking
is communicated by  gauge interactions from a ``messenger'' sector
to the squarks, slepton and Higgs. 
Since the $\mu$-parameter cannot be induced  
by  gauge interactions, one has 
$\mu=0$  
unless one enlarges the model with new interactions
\cite{us}-\cite{other}.

Property (b) leads to the  doublet-triplet splitting problem.
To embed the Higgs doublets in a complete
SU(5)-representation,  we have to 
introduce  Higgs color triplets $H_C$ and $\bar H_C$ 
such that 
${\bf \bar 5}=(\bar H_C, \bar H)$ and 
${\bf 5}=(H_C,  H)^T$.
Nevertheless, the color triplets
 cannot be light if we do not want to have a too fast proton
decay or to spoil the success of gauge coupling unification.
Thus, one needs to split the
${\bf 5}$ and ${\bf \bar 5}$ into light Higgs doublets and heavy 
color triplets.

A very attractive possibility that seems to relate 
properties (a), (b) and (c) is to assume that
the $\mu$-parameter is a dynamical variable \cite{sliding}.
In this case, its value is determined by the minimization conditions
of the potential  and one obtains that (c) leads automatically
to a doublet-triplet splitting \cite{sliding}.
To see how this works,
let us consider a SU(5)-GUT given by
\be
W=\mu\, {\bf \bar 5\, 5}+\lambda^{'}{\bf \bar 5\, 24\, 5}\, ,
\label{super}
\ee
where ${\bf 24}$ is the adjoint representation of SU(5) responsible for
the breaking of SU(5) to the SM group. 
Its VEV is assumed to be
\be
\langle{\bf 24}\rangle=M_G\, {\rm Diag}(2,2,2,-3,-3)\, ,\ \ \ M_G\simeq
10^{16}\ {\rm GeV}\, .
\label{vev}
\ee
Inserting (\ref{vev}) in (\ref{super}),
we obtain 
\be
W=(\mu+2\lambda^{'}M_G)\bar H_C H_C+
(\mu-3\lambda^{'}M_G)\bar H H\, ,
\label{super2}
\ee
and the potential for the scalar components is given by
\be
V=|\mu+2\lambda^{'}M_G|^2(|H_C|^2+|\bar H_C|^2)+
|\mu-3\lambda^{'}M_G|^2(|H|^2+|\bar H|^2)
+V_{soft}+\mbox{D-terms}\, ,
\label{pote}
\ee
where $V_{soft}$ includes the terms that softly break supersymmetry.
{}From eq.~(\ref{pote}) we can see that for values of $\mu$ 
different from $-2\lambda^{'}M_G$ or $3\lambda^{'}M_G$,
 the Higgs doublets and color 
triplets  are very heavy and  forced to get zero VEVs. 
The potential (\ref{pote}) at this minimum will then
be zero. 
On the other hand,
for   $\mu=3\lambda^{'}M_G$ the Higgs doublets are light
and their VEVs are determined by the low-energy MSSM potential.
If  at low-energies $H$ and $\bar H$
get  VEVs of order of their soft masses (of $\calo(m_Z)$),
 the potential at the minimum has a value smaller than zero.
Thus, this vacuum is energetically favored. 
The Higgs color triplets at this vacuum are very 
heavy ($M_{H_C}=5\lambda^{'}M_G$) 
in agreement with gauge coupling unification 
and proton decay limits.
There could be a third possibility with $\mu=-2\lambda^{'}M_G$
 and light 
 Higgs color triplets.
This case is however energetically disfavored because 
the  soft masses of $H_C$ and $\bar H_C$ tend to be
positive at low-energy
due to the SU(3) strong coupling (like the squark soft masses)
forcing  zero VEVs for the color triplets.

Here we will assume that $\mu$ is a dynamical variable
and calculate the  value of $\mu$ 
by minimizing  the low-energy 
effective potential  (including the soft supersymmetry 
breaking terms). 
We will show
that a local minimum exists
where the supersymmetric Higgs mass is  of $\calo(m_Z)$.
This minimum is stable under gravity corrections
if supersymmetry is broken at low-energies $\sim 10^5$ GeV.
Thus, this scenario can solve simultaneously the
doublet-triplet splitting problem and the $\mu$-problem.

\section{The dynamical value of $\mu$}

Let us promote the $\mu$-parameter   to a superfield 
\be
\mu\rightarrow\lambda S\, , 
\label{promotion}
\ee
where $S$ is a  SM singlet superfield 
 and $\lambda$ is its  coupling to $H\bar H$. 
Since we are only interested in the vacuum where the Higgs doublets
are light and the Higgs color triplets are heavy,
we expand  $\lambda S$ around  $3\lambda^{'}M_G$. This means making the 
replacement $S\rightarrow S+3\lambda^{'}M_G/\lambda$
 in the superpotential (\ref{super2}).
The low-energy 
effective potential for the neutral scalars 
 is given by,
\be
V=V_{SUSY}+V_{soft}\, ,
\label{potential}
\ee
where
\be
V_{SUSY}=|\lambda S|^2(|H|^2+|\bar H|^2)+|\lambda \bar H H|^2+
\frac{g^2+g'^2}{8}(|H|^2-|\bar H|^2)^2\, ,
\ee
and
\be
V_{soft}=m^2_{H}|H|^2+m^2_{\bar H}|\bar H|^2+m^2_S|\lambda S|^2
-(B\lambda S\bar HH+h.c.)\, .
\label{softpot}
\ee
The origin of the soft terms will be discussed in the next section.
Considering the limit $\lambda\ll 1$ (as we will see, in 
this limit  the experimental constraints are always satisfied),
we have that the potential (\ref{potential}) has a 
 stationary value for  
\bea
\frac{1}{2}m_{Z}^{2}
&=& \frac{m_{\bar H}^{2}
- m_{ H}^{2}\tan^{2}\beta}{\tan^{2}\beta - 1}-\mu^{2}\, , 
\label{min1}\\
\sin2\beta
 &=& \frac{2B\mu}
{m_{H}^{2} + m_{\bar H}^{2} + 2\mu^{2}}\, ,
\label{min2}\\
\mu&=&
\frac{m^2_WB\sin 2\beta}
{2m^2_W+g^2 m^2_S}\, .
\label{min3}
\eea
where
$m^2_W=\frac{g^2}{2}(\langle H\rangle^2+\langle \bar H\rangle^2)$,
$\tan\beta=\langle  H\rangle/\langle  \bar H\rangle$ and 
$\mu=\lambda\langle S\rangle$.
Eqs.~(\ref{min1}) and (\ref{min2}) are 
the usual minimization conditions of the MSSM. Notice that 
we have  an extra condition [eq.~(\ref{min3})] coming from
the stationarity of the potential with respect to the 
new variable $S$. 
We still have to 
guarantee that
 eqs.~(\ref{min1})-(\ref{min3}) lead to a 
(at least, local)
minimum of the potential. 
This means that
the scalar mass matrices must have  positive eigenvalues. 
While
 charged and pseudoscalar 
Higgs masses turn out to be  always positive,
we find that the condition of positive masses for the 
real part of the neutral scalars  is very restrictive.
The  sign of the  determinant  of the scalar mass matrix is given by
the quantity
\be
{\rm Det}{\cal M}^2\propto \left[1+x-\frac{x^2}{\cos^22\beta}
(1+\frac{B^2}{m^2_Z})-\frac{x^3}{\cos^22\beta}\right]\, ,
\label{deter}
\ee
where $x\equiv g^2 m^2_S/(2m^2_W)$.
Requiring
${\rm Det}{\cal M}^2>0$, we obtain  a bound
on  $x$. 
This bound is approximately given by 
\be
|x|<
\left|\frac{m_Z\cos 2\beta}{\sqrt{m^2_Z+B^2}}\right|\, .
\label{bound}
\ee
We can now
use eqs.~(\ref{min3}) and (\ref{deter}), and  infer  the values 
of $\mu$ that lead to 
${\rm Det}{\cal M}^2>0$
as a function of  $\tan\beta$ and $B$.
In Fig.~1  we plot the allowed area of the plane $\mu$--$\tan\beta$ for
different values of $B$.
This area is well approximate 
by 
values of $\mu$ inside the interval 
\footnote{Except for the region $B< m_Z$ and $\cos 2\beta\simeq -1$.
In this region, however, we find $|\mu|\lappeq 50$ GeV.}
\be
\mu=\frac{1}{2}B\sin 2\beta\left(1\pm\frac{m_Z\cos 2\beta}
{\sqrt{m^2_Z+B^2}}\right)^{-1}\, ,
\label{mui}
\ee
that can be obtained using eqs.~(\ref{min3}) and (\ref{bound}).
As  $B$ increases,  $\mu$ approaches to the central value
\be
\mu\simeq \frac{1}{2}B\sin 2\beta\, ,
\label{mu}
\ee
that we plot in Fig.~1 as a dashed line. 
This is our  prediction for $\mu$ . 
We see that in order to 
have large values of $\mu$, we need large values for $B$ 
and/or small values for $\tan\beta$.
For example, a $|\mu|\gappeq 50$ GeV such that Higgsinos 
 escape from LEP1.5 detection \cite{aleph} requires
 $\tan\beta\lappeq 3.5,5,6$
for $B\simeq 100,150,200$ GeV.

\begin{figure}[htb]
\center{\mbox{\epsfig{file=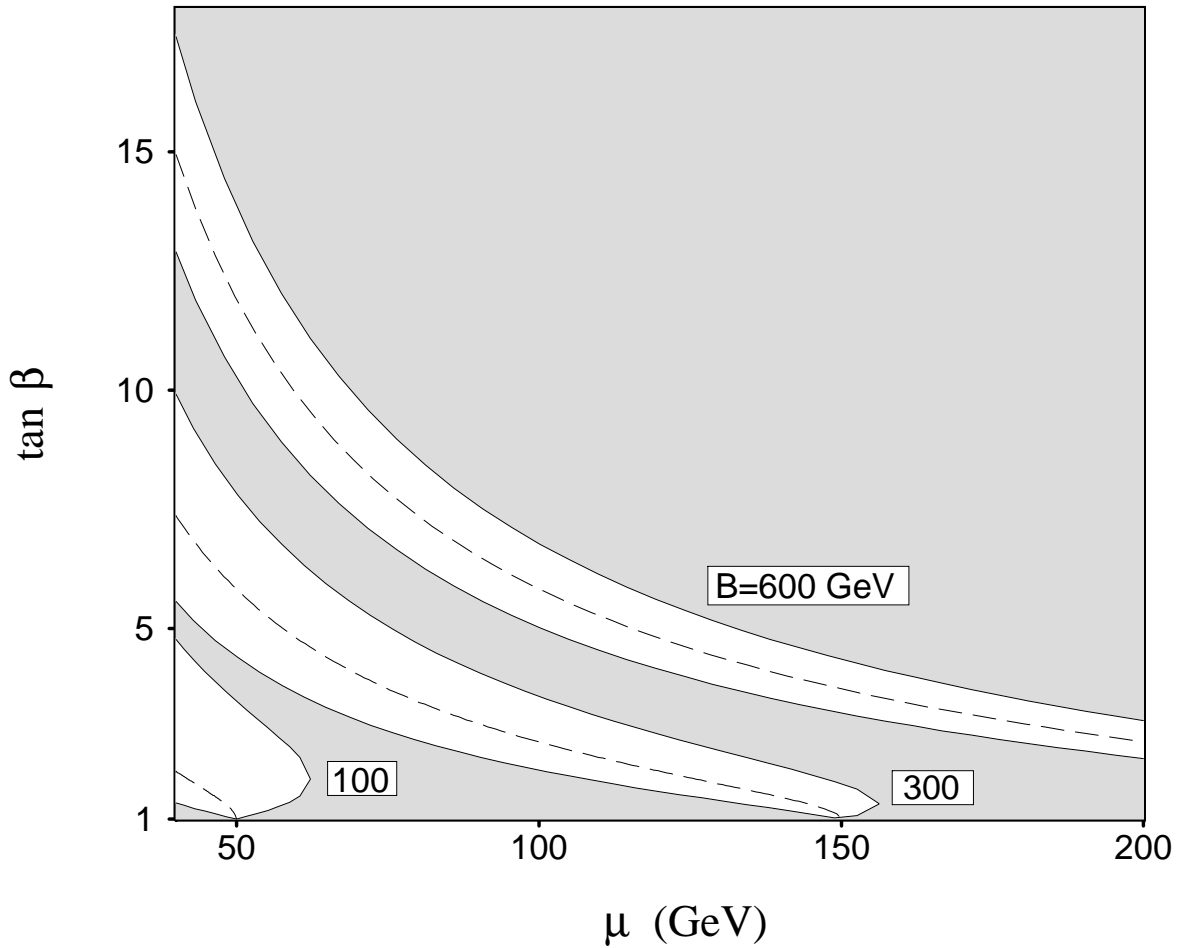,width=16.5cm}}}
\caption{Allowed region (in white) of the plane $\mu$--$\tan\beta$ for
different values of $B$. The dashed lines correspond to
$\mu=B\sin 2\beta/2$. }
\label{fig1}
\end{figure}

In the limit $m_Z\ll B$, we can use 
eqs.~(\ref{min2}) and 
(\ref{mu}) to write
$\mu$ 
as a function  of the parameters of the MSSM:
\be
\mu^2\simeq\frac{1}{2}(B^2-m^2_{H}-m^2_{\bar H})
\, .
\ee
Requiring $0\leq \sin^2 2\beta\leq 1$, we obtain 
 that $B^2$ has to lay in the window
\be
2(m^2_{H}+m^2_{\bar H}) \gappeq B^2 \gappeq(m^2_{H}+m^2_{\bar H})\, .
\ee

\section{Origin of the soft breaking terms}

Supersymmetry is usually assumed to be broken in a ``hidden'' sector.
The supersymmetry breaking is  
 transmitted from the hidden to the observable sector by 
 either gravity or
gauge interactions. In both scenarios
 soft terms  like those in 
 eq.~(\ref{softpot})  are induced and 
are proportional to 
$F/M\simeq\calo(m_Z)$ 
where $\sqrt{F}$ is the scale of supersymmetry breaking 
in the hidden sector 
 and $M$ 
is the messenger mass in  GMSB models, or the Planck mass 
($M_P$) if gravity mediates the supersymmetry breaking.
In a model with a dynamical $\mu$, however, there are  two  possible
extra soft-terms that can  be induced \cite{nilles}:
\be
V^{'}_{soft}=m^2_{12}\bar H H+\rho\lambda S+h.c.\, .
\label{extrasoft}
\ee
The origin of these terms is different from that of eq.~(\ref{softpot});
 they
turn out not to be  proportional to $F/M$ 
 and can destabilize the $m_Z-M_G$
hierarchy \cite{nilles}.
Here we will study the origin of these extra terms and the 
 constraints on the scale $\sqrt{F}$ derived from the requirement
$\rho^{1/3}, m_{12}\lappeq m_Z$.

The terms of eq.~(\ref{extrasoft}) 
can be generated from different sources depending on the underlying
theory at high energies:

\noindent a)  In supergravity theories  with flat
K{\"a}hler metric, 
there are contributions to
$\rho$ and $m_{12}$   arising
when we shift the singlet, 
$S\rightarrow S+3\lambda^{'}M_G/\lambda$ (see below eq.~(\ref{promotion})),
 in the gravity-induced soft supersymmetry breaking  terms: 
\bea
m^2_{3/2}
|S|^2&\rightarrow& 
\frac{3}{5\lambda}m^2_{3/2}M_{H_C}S+\dots\, ,\nonumber\\
m_{3/2}(\lambda[1+\epsilon]S-3\lambda^{'}M_G)\bar HH
&\rightarrow& 
\frac{3}{5}\epsilon m_{3/2}M_{H_C}\bar HH+\dots\, ,
\label{tadpole2}
\eea
where $M_{H_C}=5\lambda{'}M_G$ and 
$m_{3/2}={F}/{(\sqrt{3}M_P)}$  is the gravitino mass.
 $\epsilon$
parametrizes deviations from  proportionality between the
superpotential (\ref{super2}) and the trilinear soft terms.
Even if exact proportionality  holds 
at $M_P$
($\epsilon=0$), it will not hold
at $M_G$ due to loop effects. Thus, 
 $\epsilon\simeq 1/(4\pi)^2\simeq 10^{-2}$.
The
 stability of the weak scale requires (from eq.~(\ref{tadpole2}))
$\frac{3}{5}\epsilon m_{3/2}M_{H_C}\lappeq m^2_Z$ that 
leads, for 
$M_{H_C}\sim 10^{15}-10^{16}$ GeV, to
a  bound on $\sqrt{F}$:
\be
\sqrt{F}\lappeq 20-60\ {\rm TeV}\, .
\label{boundextra}
\ee
This constraint can be relaxed if $m_{3/2}\ll F/M_P$  like in no-scale 
models \cite{noscale}, or can
disappear if the  MSSM is not embedded in a grand-unified
theory (in such a case the singlet $S$ does not get a VEV of $\calo(M_G)$
and the contributions of eq.~(\ref{tadpole2}) do not arise).

\noindent b) In  supergravity theories with nonminimal 
K{\"a}hler metric, one can have  operators like
\be
\frac{1}{M_P}\int d^4\theta SXX^\dagger\, ,
\label{operator}
\ee
where $X$ denotes the superfield (in the hidden sector)
that breaks supersymmetry.  
Once   supersymmetry is broken, $\langle X\rangle=\theta^2F$,
the above operator  generates a tadpole contribution
given by
\be
\rho\simeq \frac{F^2}{\lambda M_P}\, .
\label{tadpole1}
\ee
Requiring $\rho\lappeq (100$ GeV$)^3$,
 we obtain a bound on the supersymmetry breaking scale:
\be
\sqrt{\frac{F}{\sqrt{\lambda}}}\lappeq 10^{6}\ {\rm GeV}\, .
\label{bound1}
\ee
The contribution to 
$m_{12}$ from the operator (\ref{operator}) is zero (unless the
scalar component of $X$ gets a VEV).

\noindent c)  There are also nongravitational contributions to 
 the tadpole term  coming
from  loops of Higgs color triplets.
These contributions 
can be understood as arising from
the operator $\frac{1}{M_{H_C}}\int d^4\theta SXX^\dagger$
induced when
the heavy Higgs color triplets 
are integrated out at the one-loop level.
This  gives
\be
\rho\simeq\frac{1}{16\pi^2}\frac{M^2m^2_{H_C}}{M_{H_C}}\, ,
\label{tadpole3}
\ee
where $m_{H_C}$ is the color triplet soft mass
and $M$ is the messenger scale. 
If we impose $\rho\lappeq (100$ GeV$)^3$, 
 we get an upper bound on $M$:
\be
M\lappeq 10^{10}\ {\rm GeV}\, ,
\label{bound2}
\ee
for $M_{H_C}\simeq 10^{16}$ GeV and $m_{H_C}\simeq 100$ GeV.
There are also  contributions to $m^2_{12}$ coming
from loops of color triplets but they are small for $M\lappeq 10^{10}$.

In models where gravity mediates
the  supersymmetry breaking 
($M\simeq M_P$ and 
$\sqrt{F}\simeq 10^{10}$ GeV) the bounds
(\ref{boundextra}), (\ref{bound1}) 
 or (\ref{bound2}) are not fulfilled
and the 
mechanism described in the previous section cannot be operative
 \cite{nilles}.
On the other hand,
in GMSB models with low-energy supersymmetry breaking
, $M\simeq\sqrt{F}\simeq 10^{5}$ GeV
\cite{dine}, these
 bounds are  satisfied. 
Furthermore, in these theories
the soft mass of $S$ is 
one-loop factor suppressed with respect to 
 the soft masses of the Higgs doublets
\be
 m^2_S\simeq\frac{1}{4\pi^2}(m^2_H+m^2_{\bar H}+B^2)
\ln\frac{m_Z}{M}\, ,
\label{masss}
\ee
and the constraint (\ref{bound})  can be also satisfied.
Nevertheless, in the minimal GMSB model
the $B$-parameter at the messenger scale
is also a  one-loop factor 
smaller than the other soft masses. 
This implies a
  small $\mu$-parameter 
(for  $B\sim 10$ GeV, we find
$\mu\lappeq 15$ GeV).
A possible way out is to  have  $\rho\simeq (100$ GeV$)^3$.
In this case
\be
\mu\simeq\frac{g^2\rho}{2m^2_W+g^2m^2_S}\, ,
\ee
and  we can have $\mu\sim 100$ GeV 
even in the minimal
GMSB model. Although this possibility could be viable, 
we do not see any reason why
 $\rho=\calo(m^3_Z)$.
A more interesting  possibility is to 
consider 
GMSB models with  messenger-matter mixing \cite{mixing}
or  with messenger-Higgs mixing \cite{us}. In these models
 a large value of
$B$  can be obtained \cite{us}. 
For example, the coupling 
$y H Q\bar D_M$ where $Q$ and $D_M$ denote  the ordinary  
quark and messenger superfield respectively, 
would generate a $B$-parameter at the one-loop level 
given by
\be
 B=\frac{3y^2}{16\pi^2}\frac{ F}{M}\, .
\ee
Surprisingly,  the contribution 
to the
soft masses of the Higgs arising from $y H Q\bar D_M$
 is comparable, for  $F/M^2\lappeq 0.1$  \cite{mixing},
to the universal two-loop contribution due to 
the cancellation of
 the leading term of   $\calo (F^2/M^2)$ \cite{us,mixing}.
In these GMSB models  $B$ comes out to be  of the same order
of the other soft masses and a 
$\mu$-parameter  from eq.~(\ref{mui}) can be larger than $50$ GeV.
Considering that 
a messenger-matter mixing can  also avoid some
cosmological problems present in  GMSB theories \cite{cosmo}, 
we find this scenario very attractive. 
This is the simplest mechanism to generate a $\mu\not=0$.

\section{The light spectrum and  fine-tuning criteria}

In the limit that $\rho$ and $m_{12}$ are smaller than the weak scale,
the potential
(\ref{potential}) has an approximate extra U(1) symmetry
under which $S$
transforms nontrivially.
There is a pseudo-Goldstone boson
associated with the spontaneous
breaking of this U(1) and its mass is given by
\be
m_{PG}\simeq \lambda\sqrt{\frac{\rho}{\mu}+\frac{m^2_W\sin 2\beta}
{\mu^2g^2}m^2_{12}}\, .
\ee
Depending on the origin of $\rho$ and $m_{12}$ [(a), (b) or (c) in 
the previous section], 
 $m_{PG}$ 
is given by
\be
m_{PG}
\sim 
\left\{
\ba{ll}
100&\left(\frac{\lambda}{0.1}\right)
\left(\frac{\sqrt{F}}{10^{5}\ {\rm GeV}}\right)\ {\rm GeV}\, ,\\
100&\sqrt{\frac{\lambda}{0.1}}
\left(\frac{F}{(10^{5}\ {\rm GeV})^2}\right)\ {\rm MeV}\, ,\\
0.1&\left(\frac{\lambda}{0.1}\right)
\left(\frac{M}{10^{5}\ {\rm GeV}}\right)\ {\rm MeV}\, ,
\ea
\right.
\label{pgbo}
\ee
where we have used 
eq.~(\ref{tadpole2}),
eq.~(\ref{tadpole1})
 and eq.~(\ref{tadpole3})   respectively.
In the first case, the pseudo-Goldstone is very heavy and
can easily escape detection
\footnote{In this case  $\lambda$ could be of $\calo(1)$.
We have checked that the effect of a $\lambda\sim 1$
is to slightly enlarge the  allowed regions of Fig.~1  for $\tan\beta$ 
 close to 1.}.
In the second and third case of eq.~(\ref{pgbo})
such a  light particle with
  axion-like couplings 
is excluded by the LEP experiment if $\lambda\sim 1$. 
Nevertheless, we have the freedom to reduce $\lambda$ 
and decouple the pseudo-Goldstone from matter
without modifying the above
prediction on $\mu$ (notice that  eqs.~(\ref{min1})-(\ref{min3}) 
do not depend on $\lambda$).
In the limit of small $\lambda$, the full supermultiplet $S$ is 
in fact light (the scalar and fermion component have masses 
$\sqrt{2}\lambda m_W/g$ and $2\lambda^2 m^2_W\sin 2\beta/(g^2\mu)$
respectively) but it is also  almost decoupled from matter.
Constraints from  $Z$-decays
require  \cite{campos}   $\lambda\lappeq 0.1$.
Searches for axion-like particles in hadron collisions \cite{hadron}
put the bound 
$\lambda\lappeq 10^{-2}$, 
but this only applies for $m_{PG}\lappeq 200$ MeV.
Astrophysical
constraints  
are more severe and  imply $\lambda\lappeq 10^{-7}$.
These, however, can be evaded
 if $m_{PG}\gappeq 1$ MeV 
that can be easily satisfied.

Let us now turn to the fine-tuning criteria.
It can be seen
from eq.~(\ref{min1}) that if the soft masses of the Higgs are
much larger than $m_Z$, the $\mu$-parameter has to be fine-tuned 
\be
\mu^2\simeq \frac{{m_{\bar H}^{2}- m_{ H}^{2}\tan^{2}\beta}}
{\tan^{2}\beta - 1}\, ,
\label{finetuning}
\ee
in order to have the right value of $m^2_Z$. 
Since $\mu$ and the soft masses are
independent parameters in the MSSM, such a 
fine-tuning is  unnatural.
The degree of fine-tuning can be 
 estimated as  \cite{ft}
\be
\Delta_{m^2_{H}}
=\frac{m^2_{ H}}{m^2_Z}
\frac{\partial m^2_Z}{\partial m^2_{ H}}
\sim
\frac{m^2_{ H}}{m^2_Z}
\, ,
\label{fine}
\ee
that can be used to put upper bounds on the soft masses \cite{ft}.
In our model 
the $\mu$-parameter is a dynamical variable 
that adjusts  itself in order to 
minimize the energy; one may then think that no fine tuning at
all is needed even if soft masses are large.
However, for $B\gg m_Z$, we see from eq.~(\ref{mui})
that $\mu$ is forced to approach to its asymptotic value
$\mu= B\sin 2\beta/2$. If this equality holds, we need to
fine-tune the potential parameters to satisfy also (\ref{min1})
and (\ref{min2}); this situation is in fact
equivalent to the MSSM one. We can quantify the amount of
fine-tuning which is needed in our model when
$B\sim m_H\gg m_Z$ by following a procedure similar
to the MSSM one. Using eqs.~(\ref{min1})-(\ref{min3}) and 
(\ref{bound}), we obtain  
\be
\Delta_{m^2_{ H}}\sim\frac{m_{ H}}{m_Z}
\, .
\ee
We see that the fine-tuning
scales
linearly with the ratio $m_H/m_Z$ instead of
quadratically as in the MSSM.  This implies less fine-tuning
 to have
the electroweak scale  smaller than the sparticle masses.
Nevertheless, we have to stress  that as $B$ increases,
we need $m^2_S$ to decrease 
(see eq.~(\ref{bound})). 
This could  be unnatural if
$m^2_S$ is tied to the  Higgs doublet
soft masses such as  in eq.~(\ref{masss}).
To address this question properly, 
one needs to specify  the details of
the mechanism that generates the soft breaking terms; this is 
beyond the scope of this paper.

\section{Conclusions}

We have proposed a scenario where the supersymmetric Higgs mass
($\mu$-parameter)  is
dynamically determined.
This has allowed to calculate $\mu$ as a 
function of the soft breaking terms
of the potential and then reduce the parameters of the MSSM. 
We have found  that $\mu$ gets a weak scale value
 close to 
$B\sin 2\beta/2$. 
Thus, this scenario provides 
a solution to
the $\mu$-problem.
If the 
MSSM is embedded in a GUT, this scenario 
solves automatically
the doublet-triplet splitting problem. 
Our
 mechanism  is operative in models with low-energy supersymmetry
breaking scale  such as  in
 GMSB theories.
In such theories
we can obtain a realistic $\mu$-parameter.
We have also shown that naturalness constraints on soft masses 
seem to be less stringent
than in the usual MSSM.

\vspace{1.cm}

It is a pleasure to thank  Gia Dvali for  discussions.
The work of one of us (P. C.) is financially supported by a
grant from INFN-Frascati-Italy.

\vskip.8cm


\begin{thebibliography}{99}

\bibitem{aleph}
The ALEPH Collaboration, hep-ex/9607009. 


\bibitem{us}
G. Dvali, G.F. Giudice and A. Pomarol, Nucl. Phys. 
{\bf B478} (1996) 31.


\bibitem{dine}

M. Dine, A.E. Nelson and Y. Shirman, Phys. Rev. 
{\bf D51} (1995) 1362;
M. Dine, A.E. Nelson, Y. Nir and Y. Shirman, 
Phys. Rev. {\bf D53} (1996) 2658.


\bibitem{mixing}
M. Dine, Y. Nir and Y. Shirman, 
Phys. Rev. {\bf D55} (1997) 1501.


\bibitem{other}
T. Yanagida, Preprint hep-ph/9701394.


\bibitem{sliding}
E. Witten, Phys. Lett. {\bf B105} (1981) 267.


\bibitem{nilles}
J. Polchinski and L. Susskind, Phys. Rev. {\bf D26} (1982) 3661;
H.P. Nilles, M. Srednicki and D. Wyler, 
Phys. Lett. {\bf B122} (1983) 337.
For a general review, see for example, H.P. Nilles,
 Phys. Rep. {\bf 110} (1984) 1.


\bibitem{noscale}
E. Cremmer, S. Ferrara, C. Kounnas and D.V.
Nanopoulos, Phys. Lett. {\bf B133} (1983) 61.


\bibitem{cosmo}
S. Dimopoulos, G.F. Giudice and A. Pomarol, 
Phys. Lett. {\bf B389} (1996) 37. 


\bibitem{campos}
F. de Campos et al., hep-ph/9405382.

\bibitem{hadron}
The CHARM Collaboration, Phys. Lett. {\bf B157} (1985) 458.



\bibitem{ft}
R. Barbieri and G.F. Giudice, Nucl. Phys. {\bf B306} (1988) 63. 


\end{thebibliography}
\end{document}